\documentstyle[epsf,aps]{revtex}
\renewcommand{\thefootnote}{\fnsymbol{footnote}}
\begin{document}
\begin{flushright}
  CERN-TH/99-166 \\
  NUC-MN-99/8-T \\ 
  TPI-MINN-99/26
\end{flushright}
\vspace*{1cm} 
\setcounter{footnote}{1}
\begin{center}
  {\Large\bf Unitarization of the BFKL Pomeron on a Nucleus} \\[1cm]
  Yuri V.\ Kovchegov \\ ~~ \\ {\it Theory Division, CERN \\ CH-1211,
    Geneva, Switzerland }\\ ~~ \\ and \\ ~~ \\ {\it School of Physics
    and Astronomy, University of Minnesota, \\ Minneapolis, MN 55455,
    USA
    \renewcommand{\thefootnote}{\fnsymbol{footnote}}\setcounter{footnote}{0}
    \footnote{Permanent address}}\\ ~~ \\ ~~ \\
\end{center}
\begin{abstract} 
  We analyze the evolution equation describing all multiple hard
  pomeron exchanges in a hadronic or nuclear structure functions that
  was proposed earlier in \cite{me}. We construct a perturbation
  series providing us with an exact solution to the equation outside
  of the saturation region. The series demonstrates how at moderately
  high energies the corrections to the single BFKL pomeron exchange
  contribution which are due to the multiple pomeron exchanges start
  unitarizing total deep inelastic scattering cross section. We show
  that as energy increases the scattering cross section of the
  quark--antiquark pair of a fixed transverse separation on a hadron
  or nucleus given by the solution of our equation inside of the
  saturation region unitarizes and becomes independent of energy. The
  corresponding $F_2$ structure function also unitarizes and becomes
  linearly proportional to $\ln s$. We also discuss possible
  applications of the developed technique to diffraction.
\end{abstract}

\newcommand{\stackeven}[2]{{{}_{\displaystyle{#1}}\atop\displaystyle{#2}}}
\newcommand{\lsim}{\stackeven{<}{\sim}}
\newcommand{\gsim}{\stackeven{>}{\sim}}

\section{Introduction}

The Balitsky, Fadin, Kuraev and Lipatov (BFKL) \cite{EAK,Yay} equation
resums all leading logarithms of Bjorken $x$ for hadronic cross
sections and structure functions. The solution of the BFKL equation
grows like a power of center of mass energy $s$, therefore violating
the unitarity bound at very high energies \cite{Fro,Mar}. This is one
of the major problems of small $x$ physics, since the Froissart bound
\cite{Fro,Mar} states that the total cross section should not raise
faster than $\ln^2 s$ at asymptotically high energies. There is a
general belief that the unitarity problem could be cured by resumming
all multiple BFKL pomeron exchanges in the total cross section, or,
equivalently, in the structure function.

As was argued by Mueller in \cite{Mueller5} multiple pomeron exchanges
become important at the values of rapidity of the order of 
\begin{equation}\label{yu}
  Y_U \sim \frac{1}{\alpha_P - 1} \, \ln \frac{1}{\alpha^2},
\end{equation}
with $\alpha$ the strong coupling constant, which is assumed to be
small, and $\alpha_P - 1 = \frac{4 \alpha N_c}{\pi} \ln 2$. This
result could be obtained if one notes that one pomeron contribution to
the total cross section of, for instance, onium--onium scattering, is
parametrically of the order of $\alpha^2 \exp [(\alpha_P - 1) Y]$ and
the contribution of the double pomeron exchange is $\alpha^4 \exp [2
(\alpha_P - 1) Y]$. Since multiple pomeron exchanges become important
when the single and double pomeron exchange contributions become
comparable, we recover Eq. (\ref{yu}) by just equating the two
expressions.

After completion of the calculation of the next--to--leading order
corrections to the kernel of the BFKL equation (NLO BFKL) by Fadin and
Lipatov \cite{VSF}, and, independently, by Camici and Ciafaloni
\cite{Cia}, it was shown in \cite{Regge} that due to the running
coupling effects these corrections become important at the rapidities
of the order of
\begin{equation}\label{rc}
Y_{NLO} \sim \frac{1}{\alpha^{5/3}}.
\end{equation}
One can see that $Y_U \ll Y_{NLO}$ for parametrically small $\alpha$.
That implies that the center of mass energy at which the multiple
pomeron exchanges become important is much smaller, and, therefore, is
easier to achieve, than the energy at which NLO corrections start
playing an important role. That is multiple pomeron exchanges are
probably more relevant than NLO BFKL to the description of current
experiments \cite{Cald}.  Also multiple pomeron exchanges are much
more likely to unitarize the total hadronic cross section. In this
paper we are going to propose a solution to the problem of resummation
of multiple pomeron exchanges and show how the BFKL pomeron
unitarizes.

\begin{figure}
\begin{center}
  \epsfxsize=10cm
\leavevmode
\hbox{ \epsffile{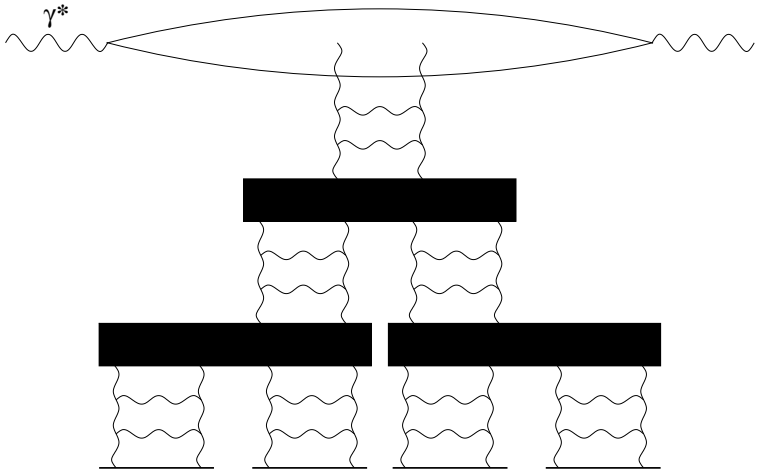}}
\end{center}
\caption{An example of the pomeron ``fan'' diagram.}
\label{fan}
\end{figure}

In the previous paper on the subject \cite{me} we proposed an equation
which resums all multiple BFKL pomeron \cite{EAK,Yay} exchanges in the
nuclear structure function. In deriving the equation we used the
techniques of Mueller's dipole model
\cite{Mueller1,Mueller2,Mueller3,MZ}. As was proven some time ago
\cite{Mueller1,Samuel} the dipole model can provide us with the
equation which is exactly equivalent to the BFKL equation
\cite{EAK,Yay}. In \cite{me} it was shown that the dipole model
methods, when applied to hadronic or nuclear structure functions,
yield us with the equation which in the traditional Feynman diagram
language resums the so--called pomeron ``fan'' diagrams in the leading
logarithmic approximation, an example of which is depicted in Fig.
\ref{fan} for the case of deep inelastic scattering.

\begin{figure}
\begin{center}
\epsfxsize=15cm
\leavevmode
\hbox{ \epsffile{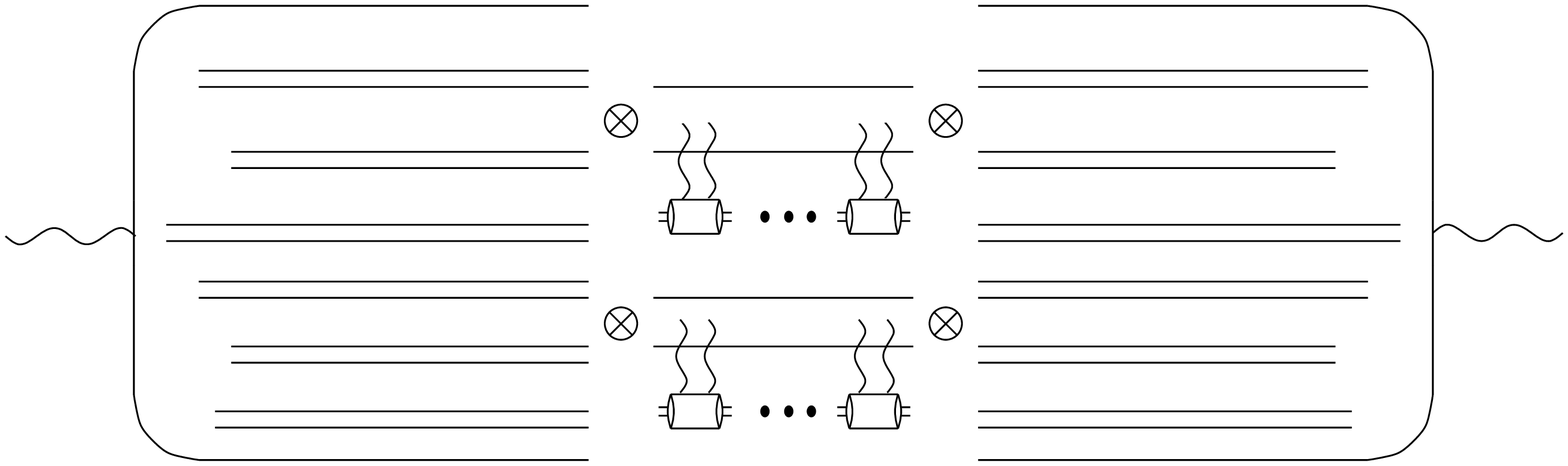}}
\end{center}
\caption{Dipole evolution in the deep inelastic scattering process as
  pictured in Ref.[1]. The incoming virtual photon develops a system
  of color dipoles, each of which rescatters on the nucleus (only two
  are shown). }
\label{dis}
\end{figure}

To review the results of \cite{me} let us consider a deep inelastic
scattering (DIS) of a virtual photon on a hadron or a nucleus. An
incoming photon splits into a quark--antiquark pair and then the
$q{\overline q}$ pair rescatters on the target hadron. In the rest
frame of the hadron all QCD evolution should be included in the wave
function of the incoming photon. That way the incoming photon develops
a cascade of gluons, which then scatter on the hadron at rest. In
\cite{me} this gluon cascade was taken in the leading longitudinal
logarithmic ($\ln 1/x$) approximation and in the large $N_c$ limit.
This is exactly the type of cascade described by Mueller's dipole
model \cite{Mueller1,Mueller2,Mueller3,MZ}. In the large $N_c$ limit
the incoming gluon develops a system of color dipoles and each of them
independently rescatters on the hadron (nucleus) \cite{me}. The
forward amplitude of the process is shown in Fig. \ref{dis}. The
double lines in Fig. \ref{dis} correspond to gluons in large $N_c$
approximation being represented as consisting of a quark and an
antiquark of different colors. The color dipoles are formed by a quark
from one gluon and an antiquark from another gluon. In Fig.  \ref{dis}
each dipole, that was developed through the QCD evolution, later
interacts with the nucleus by a series of Glauber--type multiple
rescatterings on the nucleons.  The assumption about the type of
interaction of the dipoles with the hadron or nucleus is not important
for the evolution. It could also be just two gluon exchanges. The
important assumption is that each dipole interacts with the target
independently of the other dipoles, which is done in the spirit of the
large $N_c$ limit.

For the case of a large nucleus independent dipole interactions with
the target nucleus are enhanced by some factors of the atomic number
$A$ of the nucleus compared to the case when several dipoles interact
with the same nucleon \cite{me}. This allowed us to assume that the
dipoles interact with the nucleus independently. The summation of
these contribution corresponds to summation of the pomeron ``fan''
diagrams of Fig. \ref{fan}. However, in general there is another class
of diagrams, which we will call pomeron loop diagrams.  In a pomeron
loop diagram a pomeron first splits into two pomerons, just like in
the fan diagram, but then the two pomerons merge back into one
pomeron. In other words the graphs containing pomerons not only
splitting, but also merging together will be referred to as pomeron
loop diagrams. In a large nucleus these pomeron loop diagrams could be
considered small, since they would be suppressed by powers of $A$.
This approximation strictly speaking is not valid for the calculation
of a hadronic wave function. At very high energies it also becomes not
very well justified even in the nuclear case for the following reason.
The contribution of each additional pomeron in a fan diagram on a
nucleus is parametrically of the order of $\alpha^2 A^{1/3}
e^{(\alpha_P - 1)Y}$, which is large at the rapidities of the order of
$Y_U$ given by Eq. (\ref{yu}). The contribution of an extra pomeron in
a pomeron loop diagram is of the order of $\alpha^2 e^{(\alpha_P -
  1)Y}$, which is not enhanced by powers of $A$ and is, therefore,
suppressed in the nuclear case. However, one can easily see that at
extremely high energies, when $\alpha^2 e^{(\alpha_P - 1)Y}$ becomes
greater than or of the order of one, pomeron loop diagrams become
important even for a nucleus, still being smaller than the fan
diagrams.

Resummation of pomeron loop diagrams in a dipole wave function seems
to be a very hard technical problem, possibly involving NLO BFKL or,
even, next-to-next-to-leading order BFKL kernel calculations
\cite{KMW}.  Nevertheless, it is the author's belief that these
pomeron loop diagrams are not going to significantly change the high
energy behavior of the structure functions.  Still in the rigorous
sense the results of this paper should be considered as resumming the
fan diagrams only, which is strictly justified only for a large
nucleus.

In \cite{me}, rewriting the $F_2$ structure function of the target as
a convolution of the wave function of the $q \overline{q}$
fluctuations of a virtual photon with the propagator of the
quark--antiquark pair through the nucleus, we obtained
\begin{equation}\label{f2}
  F_2 (x, Q^2) = \frac{Q^2}{4 \pi^2 \alpha_{EM}} \int \frac{d^2 {\bf
      x}_{01} d z }{2 \pi} \, \Phi ({\bf x}_{01},z) \ d^2 b_0 \ 
  N({\bf x}_{01},{\bf b}_0 , Y) ,
\end{equation}
where the photon's wave function 
\begin{equation}
\Phi ({\bf x}_{01},z) = \Phi_T ({\bf x}_{01},z) + \Phi_L ({\bf x}_{01},z)
\end{equation}
consists of transverse
\begin{mathletters}
\begin{equation}\label{ft}
  \Phi_T ({\bf x}_{01},z) = \frac{2 N_c \alpha_{EM}}{\pi} a^2 \ K_1^2
  (x_{01} a) \ [z^2 + (1 - z)^2] 
\end{equation}
and longitudinal 
\begin{equation}\label{fl}
  \Phi_L ({\bf x}_{01},z) = \frac{2 N_c \alpha_{EM}}{\pi} \ 4 Q^2 z^2
  (1 - z)^2 \ K_0^2 (x_{01} a)
\end{equation}
\end{mathletters}
contributions, with $a^2 = Q^2 z (1 - z)$, and we assumed that quarks
are massless and have only one flavor and unit electric charge.  As
always $Q^2$ is the virtuality of the photon and $x$ is the Bjorken
$x$-variable. $z$ is the fraction of the photons light cone momentum
carried by the quark, ${\bf x}_{01} = {\bf x}_1 - {\bf x}_0$ is the
transverse separation of the quark--antiquark pair, ${\bf b}_0$ is the
impact parameter of the original pair and $Y \sim \ln (1/x)$ is the
rapidity variable.  $N({\bf x}_{01},{\bf b}_0 , Y)$ is the propagator
of the quark--antiquark pair through the nucleus, or, more correctly,
the forward scattering amplitude of the $q \overline{q}$ pair with the
nucleus (hadron). It obeys the equation \cite{me}
\begin{eqnarray*}
  N({\bf x}_{01},{\bf b}_0, Y) = - \gamma ({\bf x}_{01},{\bf b}_0) \,
  \exp \left[ - \frac{4 \alpha C_F}{\pi} \ln \left(
      \frac{x_{01}}{\rho} \right) Y \right] + \frac{\alpha C_F}{\pi^2}
  \int_0^Y d y \, \exp \left[ - \frac{4 \alpha C_F}{\pi} \ln \left(
      \frac{x_{01}}{\rho} \right) (Y - y) \right]
\end{eqnarray*}  
\begin{eqnarray}\label{eqN}
  \times \int_\rho d^2 x_2 \frac{x_{01}^2}{x_{02}^2 x_{12}^2} \, [ 2
  \, N({\bf x}_{02},{\bf b}_0 + \frac{1}{2} {\bf x}_{12}, y) - N({\bf
    x}_{02},{\bf b}_0 + \frac{1}{2} {\bf x}_{12}, y) \, N({\bf
    x}_{12},{\bf b}_0 + \frac{1}{2} {\bf x}_{02}, y) ] ,
\end{eqnarray}
with $\gamma ({\bf x}_{01},{\bf b}_0)$ the propagator of a dipole of
size ${\bf x}_{01}$ at the impact parameter ${\bf b}_0$ through the
target nucleus or hadron, which was taken to be of Glauber form in
\cite{me}:
\begin{equation}\label{gla}
   \gamma ({\bf x}_{01},{\bf b}_0) = \exp \left[ -
       \frac{\alpha \pi^2}{2 N_c S_\perp} {\bf x}_{01}^2 A xG ( x
       ,1/{\bf x}_{01}^2 ) \right] - 1.
\end{equation}
here $S_\perp$ is the transverse area of the hadron or nucleus, $A$ is
the atomic number of the nucleus, and $xG$ is the gluon distribution
of the nucleons in the nucleus, which was taken at the two gluon
(lowest in $\alpha$) level \cite{Mueller4}. Eq. (\ref{gla}) is written
for a cylindrical hadron or nucleus. In the large $N_c$ limit $C_F
\approx N_c /2$. In Eq. (\ref{eqN}) $\rho$ is an ultraviolet regulator
\cite{me,Mueller1}.

Eq. (\ref{eqN}) resums all multiple hard pomeron exchanges in the
$F_2$ structure function. The linear in $N$ term on the right hand
side of Eq. (\ref{eqN}) corresponds to the usual BFKL evolution
\cite{Mueller1}, whereas the quadratic term in $N$ is in a certain way
responsible for triple pomeron vertices. This quadratic term on the
right hand side of Eq. (\ref{eqN}) introduces some damping effects,
which at the end unitarize the growth of the BFKL pomeron. We note
that an equation similar to (\ref{eqN}) was suggested by Gribov, Levin
and Ryskin (GLR) in momentum space representation in \cite{GLR} and
was subsequently studied by Bartels and Levin in \cite{BL}. However,
usually by the GLR equation one implies an equation summing all fan
diagrams in the double leading logarithmic approximation. Eq.
(\ref{eqN}) resums all the fan diagram in the leading longitudinal
logarithmic approximation.  In the double logarithmic limit (large
$Q^2$) Eq. (\ref{eqN}) reduces to the GLR equation. We also note that
an equation very similar to Eq. (\ref{eqN}) has been proposed by
Balitsky in \cite{Bali} to be used in describing the wavefunctions of
each of the quarkonia in onium--onium scattering, similarly to
\cite{Mueller2,NP}.  A renormalization group techniques based on
McLerran--Venugopalan \cite{MV} model have been developed by
Jalilian-Marian, Kovner, Leonidov, and Weigert \cite{JKLW} leading to
an equation which is also supposed to resum multiple hard pomeron
exchanges \cite{JKLW}.  Unfortunately the resulting equation
\cite{JKLW} differs from our Eq.  (\ref{eqN}). An equation summing all
multiple pomeron diagrams in the double logarithmic limit has been
proposed by Ayala, Gay Ducati and Levin \cite{AGL}, which is also
different from the appropriate limit of our Eq. (\ref{eqN}).

In this paper we are going to analyze Eq. (\ref{eqN}). In Sect. II we
will find a solution of Eq. (\ref{eqN}) in the form of a series in Eq.
(\ref{psol}) which will be constructed by solving Eq. (\ref{eqN})
perturbatively in the momentum space. The details of transforming Eq.
(\ref{eqN}) into momentum space will be given in the Appendix A with
the result given by Eq. (\ref{bsp}).  The terms of the series will
correspond to multiple pomeron exchange contributions. Partial sums of
the series will be plotted in Fig. \ref{uni}, which shows how multiple
pomeron exchanges unitarize the amplitude.  We will argue that the
kinematic region for which the series is convergent corresponds to the
case when the DIS cross sections and structure functions have not
reached saturation. In a certain sense we will define the saturation
region as the kinematic regime where the series ceases to converge. By
resumming the terms of the series corresponding to exchanges of large
numbers of pomerons we construct a function which we can analytically
continue outside the region of convergence of the series of Eq.
(\ref{psol}). That would provide us with the ansatz for the solution
of Eq. (\ref{bsp}) outside of the saturation region [see Eq.
(\ref{asym})]. We will prove that this ansatz actually is the only
possible energy independent solution of Eq. (\ref{bsp}).

To transform our results into coordinate space on would have to
separate the momentum integration in two regions, one of them being
above and the other one being below the saturation scale. Since we
will find only the asymptotic solution inside the saturation region
our results may not be very precise in determining the coordinate
space amplitude. A numerical solution of Eq. (\ref{eqN}) would
probably allow one to achieve better precision. Nevertheless for
asymptotically high energies we can transform the amplitude into
coordinate space obtaining Eq. (\ref{nsa}). Thus we will show that
multiple pomeron exchanges unitarize the forward amplitude, making it
independent of energy at high energies, which is shown in Fig.
\ref{nsat}.

In Sect. III we will analyze the high energy asymptotics of the $F_2$
structure function. We will show that, contrary to some expectations,
$F_2$ structure function does not become independent of energy at high
energies. It becomes linearly proportional to $\ln s$, still
satisfying the Froissart bound [see Eq. (\ref{f2a})].

In Sect. IV we will discuss how in the framework of the dipole model
one can describe diffractive scattering by including the multiple
pomeron evolution in rapidity of Eq. (\ref{eqN}) in the quasi--elastic
structure functions considered in \cite{mL}. We will derive an
expression for the quasi--elastic structure function $F_2^D$
(\ref{f2diff}) and discuss its asymptotics (\ref{f2da}). Finally, we
will conclude in Sect. V by summarizing our results and discussing the
possible implications of the developed techniques on different issues
of small-$x$ physics.

\section{Perturbative solution of the evolution equation}

We start by considering a DIS on a large cylindrical nucleus. This
assumption about the size and geometry of the nucleus is being done
for simplicity of calculations and our results could be easily
generalized to a large nucleus of any shape. If the nucleus is
sufficiently large we can assume that the forward amplitude of the
scattering of the original quark--antiquark pair on the nucleus,
$N({\bf x}_{01},{\bf b}_0, Y)$, is a slowly varying function of the
impact parameter of the virtual photon ${\bf b}_0$. This is equivalent
to assuming that the transverse extent of the dipole wave function is
much smaller than the size of the nucleus, and that the impact
parameter between the virtual photon and the nucleus is smaller than
the radius of the nucleus $R$. We note that in order to construct
$N({\bf x}_{01},{\bf b}_0, Y)$ in \cite{me} we integrated over the
impact parameters ${\bf b}_i$'s of the nucleons with which each of the
pomerons in Fig.  \ref{fan} interacted. That way ${\bf b}_0$ does not
have a meaning of the impact parameter of a pomeron exchange anymore
and ${\bf b}_0$ dependence in $N({\bf x}_{01},{\bf b}_0, Y)$ becomes
purely ``geometrical'' .  In a more formal language we assume that
$x_\perp \ll R$, where $x_\perp$ is the typical transverse size of the
dipole wave function, and $b_0 < R$ ($b_0$ is zero at the center of
the nucleus).  If these conditions are satisfied we may neglect ${\bf
  x}_{12}$ and ${\bf x}_{02}$ compared to ${\bf b}_0$ on the right
hand side of Eq. (\ref{eqN}), which means that the changes in $N({\bf
  x}_{01},{\bf b}_0, Y)$ when ${\bf b}_0$ is varied by ${\bf x}_\perp$
are negligibly small. Thus suppressing the ${\bf b}_0$ dependence we
write $N({\bf x}_{01}, Y)$ instead of $N({\bf x}_{01},{\bf b}_0, Y)$
in Eq.(\ref{eqN}).  Then differentiating Eq.  (\ref{eqN}) with respect
to $Y$ we can rewrite it as
\begin{eqnarray*}
  \frac{\partial N({\bf x}_{01}, Y)}{\partial Y} = \frac{2 \alpha
    C_F}{\pi^2} \, \int_\rho d^2 x_2 \left[ \frac{x^2_{01}}{x^2_{02}
      x^2_{12}} - 2 \pi \delta^2 ({\bf x}_{01} -{\bf x}_{02}) \ln
    \left( \frac{x_{01}}{\rho} \right) \right] N({\bf x}_{02}, Y)
\end{eqnarray*}
\begin{eqnarray}\label{1diff}
  - \frac{\alpha C_F}{\pi^2} \, \int d^2 x_2 \,
  \frac{x^2_{01}}{x^2_{02} x^2_{12}} \, N({\bf x}_{02}, Y) \, N({\bf
    x}_{12}, Y),
\end{eqnarray}
with $N({\bf x}_{01}, 0) = - \gamma( {\bf x}_{01})$ as the initial
condition, where we also suppressed (but did not neglect) the impact
parameter dependence of $\gamma$. Now one can see that the evolution
equation (\ref{1diff}) itself in the large nucleus approximation does
not impose any impact parameter ${\bf b}_0$ dependence.  The latter is
only included in the initial conditions given by $\gamma( {\bf
  x}_{01}, {\bf b}_0)$. Since $\gamma$ is a slowly varying function of
${\bf b}_0$ (constant for a cylindrical nucleus) our assumption about
the slow dependence of $N$ on ${\bf b}_0$ is self consistent.
Physically this implies that before the collision, when the incoming
virtual photon develops its dipole evolved wave function, it does not
have any information about the nucleus. The information about the
thickness of the nuclear medium as a function of the impact parameter
comes through $\gamma$ when the dipoles interact with the nucleus.

Let us rewrite
\begin{mathletters}
\begin{eqnarray}\label{trfm}
  N(x_\perp, Y) = x_\perp^2 \int \frac{d^2 k}{2 \pi} \, e^{i {\bf k}
    \cdot {\bf x}} \, {\tilde N} (k, Y) = x_\perp^2 \int_0^\infty dk
  \, k \, J_0 (k x_\perp) \, {\tilde N} (k, Y).
\end{eqnarray}
We note that the inverse of this transformation is
\begin{eqnarray}\label{invtrfm}
  {\tilde N} (k, Y) = \int \frac{d^2 x}{2 \pi x_\perp^2} \, e^{- i
    {\bf k} \cdot {\bf x}} \, N(x_\perp, Y) = \int_0^\infty \frac{d
    x_\perp}{x_\perp} \, J_0 (k x_\perp) \, N(x_\perp, Y).
\end{eqnarray}
\end{mathletters}
Since $N(x_\perp, Y)$ does not depend on the direction of ${\bf
  x}_\perp$, ${\tilde N} (k, Y)$ does not depend on the direction of
the relative transverse momentum of the $q \overline{q}$ pair ${\bf
  k}_\perp$. That allowed us to simplify the integrations in Eqs.
(\ref{trfm}) and (\ref{invtrfm}). Now Eq.  (\ref{1diff}) becomes
\begin{eqnarray}\label{bsp}
  \frac{\partial {\tilde N} (k, Y)}{ \partial Y} = \frac{2 \alpha
    N_c}{\pi} \, \chi \left( - \frac{\partial}{\partial \ln k} \right)
  \, {\tilde N} (k, Y) - \frac{\alpha N_c}{\pi} \, {\tilde N}^2 (k,
  Y),
\end{eqnarray}
where
\begin{equation}\label{chi}
  \chi (\lambda) = \psi (1) - \frac{1}{2} \psi \left( 1 -
    \frac{\lambda}{2} \right) - \frac{1}{2} \psi \left(
    \frac{\lambda}{2} \right)
\end{equation}
is the eigenvalue of the BFKL kernel \cite{EAK,Yay,Mueller1} with
$\psi (\lambda) = \Gamma ' (\lambda) / \Gamma (\lambda)$. In Eq.
(\ref{bsp}) the function $\chi (\lambda)$ is taken as a differential
operator with $\lambda = - \partial / \partial \ln k$ acting on
${\tilde N} (k, Y)$. We also put $N_c / 2$ instead of $C_F$ everywhere
in the spirit of the large $N_c$ approximation. The details of
obtaining Eq. (\ref{bsp}) from Eq.  (\ref{1diff}) are presented in the
Appendix A. Eq. (\ref{bsp}) shows explicitly that the first, linear,
term on its right hand side gives the usual BFKL equation, whereas the
quadratic term corresponds to the triple pomeron vertex, which
unitarizes the growth of the BFKL pomeron.

One can easily prove that Eq.  (\ref{bsp}) has a unique solution
satisfying given initial condition at $Y=0$. The strategy of the proof
is conventional: assume that there are two different solutions of Eq.
(\ref{bsp}) satisfying given initial conditions. Then one can easily
derive an equation for the function given by the difference of these
solutions with the initial condition stating that the function is
equal to zero when $Y=0$. After a little analysis of the resulting
equation with that initial condition one could easily see that the
difference between the two solutions of the original Eq.  (\ref{bsp})
is zero. That way one proves uniqueness of the solution of Eq.
(\ref{bsp}).

We want to find a solution of Eq. (\ref{bsp}) satisfying the initial
condition given by the BFKL pomeron contribution at relatively
``small'' rapidity $Y \sim 1/\alpha$. Unfortunately finding an exact
analytical solution of Eq. (\ref{bsp}) seems to be a very difficult
task, partly because Eq. (\ref{bsp}) is non--linear, partly because of
a complicated structure of the BFKL kernel $\chi (\lambda)$. Instead
we are going to construct a series resulting from the perturbative
solution of Eq.  (\ref{bsp}), which, in the region where it is
convergent, provides us with the {\it exact} solution of Eq.
(\ref{bsp}). As we will show below the region of convergence of that
series corresponds to the regime when the DIS cross sections and
structure functions are {\it not} saturated. In the saturation regime
the series diverges, but allows us to construct an asymptotic solution
by analytical continuation.

Let us start by assuming that at $Y \sim 1/\alpha$ the function
$\tilde{N}$ is very small, $ \tilde{N} \ll 1$. This allows us to
neglect the quadratic term in Eq. (\ref{bsp}), since it would be of
higher order in $\tilde{N}$. Then Eq. (\ref{bsp}) becomes
\begin{eqnarray}\label{bfkl}
  \frac{\partial {\tilde N}_1 (k, Y)}{ \partial Y} = \frac{2 \alpha
    N_c}{\pi} \, \chi \left( - \frac{\partial}{\partial \ln k} \right)
  \, {\tilde N}_1 (k, Y),
\end{eqnarray}
which corresponds to the usual BFKL equation \cite{EAK,Yay}. The
solution of Eq. (\ref{bfkl}) is given by 
\begin{equation}\label{solb}
{\tilde N}_1 (k, Y) = \exp \left[ \frac{2 \alpha N_c}{\pi} \, Y \, \chi
\left( - \frac{\partial}{\partial \ln k} \right) \right] \, C (k),
\end{equation}
with $C (k)$ being some unknown function of $k$, to be specified by
initial conditions.

 Rewrite  $C (k)$ as a Mellin transformation
\begin{equation}\label{cmel}
  C (k) = \int \frac{d \lambda}{2 \pi i} \left( \frac{k}{\Lambda}
  \right)^\lambda C_\lambda  ,
\end{equation}
where $\Lambda \sim \Lambda_{QCD} $ is some scale characterizing the
nucleus or hadron. Here the integration in $\lambda$ runs parallel to
the imaginary axis to the right of all the singularities of
$C_\lambda$.  Substituting Eq. (\ref{cmel}) into Eq.  (\ref{solb}) we
obtain
\begin{equation}\label{nmel}
{\tilde N}_1 (k, Y) = \int \frac{d \lambda}{2 \pi i} \, \exp \left[ \frac{2
\alpha N_c}{\pi} \, Y \, \chi \left( - \lambda \right) 
\right] \left( \frac{k}{\Lambda} \right)^\lambda C_\lambda  .
\end{equation}

Assuming that the transverse momentum of the $q \overline{q}$ pair $k$
is not very large, such that $\ln (k/\Lambda) \ll \alpha Y N_c$ and
that $C_\lambda (Y)$ is a slowly varying function of $\lambda$ we come
to the conclusion that the integral in Eq. (\ref{nmel}) is dominated
by the saddle point in the exponential in the vicinity of $\lambda = -
1$ in $\chi (- \lambda)$.  Expanding $\chi (- \lambda) \approx 2 \ln 2
+ \frac{7}{4} \zeta (3) ( \lambda + 1 )^2$, with $\zeta (z)$ the
Riemann zeta function, we can perform a saddle point approximation in
the integral of Eq.  (\ref{nmel}). The saddle point is given by
\begin{equation}\label{sp}
\lambda_{sp} = -1 - a (k,Y), 
\end{equation}
where we have defined
\begin{equation}\label{a}
a (k,Y) = \frac{\pi \ln (k/\Lambda)}{7 \alpha N_c \zeta (3) Y},
\end{equation}
which is small. The result of the integration yields
\begin{equation}\label{n1}
{\tilde N}_1 (k, Y) = C_{-1} \, \frac{\Lambda}{k} \,
\frac{\exp [ (\alpha_P - 1)Y] }{\sqrt{14 \alpha N_c \zeta (3) Y}} 
\exp \left( - \frac{\pi}{14 \alpha N_c \zeta (3) Y} \ln^2 
\frac{k}{\Lambda} \right) \equiv P_1 (k, Y),
\end{equation}
a factor usually associated with the zero momentum transfer single
hard pomeron exchange, which we denoted by $P_1$. $C_{-1}$ is
determined from initial conditions. If the initial conditions are
given by the two gluon exchange approximation, then $C_{-1} \sim
\alpha^2$.  That way, parametrically, when $Y \sim 1/\alpha$, the
amplitude $\tilde{N}_1 \sim \alpha^2 \ll 1$, and our assumption of the
smallness of $\tilde{N}$ at $Y \sim 1/\alpha$ is justified.

Eq. (\ref{n1}) provides us with the usual single BFKL pomeron
solution. Now we are going to find corrections to this expression
resulting from Eq. (\ref{bsp}), which will form a series
\begin{equation}\label{ps}
\tilde{N} (k, Y) = \sum_{n=1}^\infty {\tilde N}_n (k, Y),
\end{equation}
with ${\tilde N}_n$ being of the order of ${\tilde N}_1^n$. The first
correction is ${\tilde N}_2$. Rewriting ${\tilde N} = {\tilde N}_1 +
{\tilde N}_2$, substituting it back into Eq. (\ref{bsp}), neglecting
non--linear terms in ${\tilde N}_2$ and employing Eq. (\ref{bfkl}) we
obtain
\begin{equation}\label{n2eq}
\frac{\partial {\tilde N}_2 (k, Y)}{ \partial Y} = \frac{2 \alpha
    N_c}{\pi} \, \chi \left( - \frac{\partial}{\partial \ln k} \right)
  \, {\tilde N}_2 (k, Y) - \frac{\alpha N_c}{\pi} \, {\tilde N}_1 (k,
  Y)^2,
\end{equation}
where ${\tilde N}_1$ is given by Eq. (\ref{nmel}). One can easily see
that the solution of Eq. (\ref{n2eq}) with the initial condition that
it disappears at small $Y$ is
\begin{equation}\label{n21}
{\tilde N}_2 (k, Y) = - \frac{\alpha N_c}{\pi} \int_0^Y dy \exp \left[ 
\frac{2 \alpha N_c}{\pi} \, (Y - y) \,\chi \left( - \frac{\partial}
{\partial \ln k} \right)\right] [{\tilde N}_1 (k, y)^2 ].
\end{equation}
Plugging ${\tilde N}_1$ of Eq. (\ref{nmel}) into Eq. (\ref{n21}) and
integrating over $y$ yields
\begin{eqnarray*}
{\tilde N}_2 (k, Y) = - \int \frac{d \lambda_1 
\, d \lambda_2}{(2 \pi i)^2} \, C_{\lambda_1} \, C_{\lambda_2} \, 
\left(\frac{k}{\Lambda}\right)^{\lambda_1 + \lambda_2} \, \frac{1}{2 
[\chi (- \lambda_1) + \chi (- \lambda_2) - \chi (- \lambda_1 - 
\lambda_2)]} 
\end{eqnarray*}
\begin{eqnarray}\label{n22}
\times \left[ \exp \left( 2 \, \frac{\alpha N_c}{\pi} \, Y \, [\chi (- \lambda_1) 
+ \chi (- \lambda_2)] \right) - \exp \left( 2 \, \frac{\alpha N_c}{\pi} \, Y \,
\chi (- \lambda_1 - \lambda_2) \right)  \right] .
\end{eqnarray}
Eq. (\ref{n2eq}) corresponds to putting one triple pomeron vertex
(given by quadratic term in ${\tilde N}$) into the DIS diagram, so
that ${\tilde N}_2$ consists of one pomeron splitting into two.  The
integration over $y$ in Eq. (\ref{n21}) corresponds to integration
over all possible rapidity positions of the triple pomeron vertex
between the incoming quark--antiquark pair and the target. Thus the
first exponential term in the square brackets in Eq. (\ref{n22})
corresponds to the case when the pomeron splitting occurs very close
(in rapidity) to the $q \overline{q}$ pair, so that most of the
rapidity interval is covered by two pomeron contribution. The second
exponential term in Eq. (\ref{n22}) corresponds to the situation when
the splitting happens very close to the target, so that most of the
interaction corresponds to the single pomeron exchange. Therefore the
first exponential term in Eq. (\ref{n22}) corresponds to two pomeron
contribution and the second term corresponds to the one pomeron
contribution. Similar conclusions have been reached by Navelet and
Peschanski in \cite{NP} when analyzing two pomeron contribution to
onium--onium scattering in dipole model \cite{Mueller2}. As we will
see below the two pomeron contribution gives a term proportional to
$\exp [2 (\alpha_P - 1) Y]$, while the one pomeron exchange is
proportional to $\exp [(\alpha_P - 1) Y]$. That way the two pomeron
term dominates in ${\tilde N}_2$ and the single pomeron term could be
safely neglected.  One should also remember that ${\tilde N}_2$ is
proportional to the square of $C_\lambda$, so that the single pomeron
exchange term from ${\tilde N}_2$ represents just an order $\alpha^2$
(i.e., next-to-next-to-leading order) correction to ${\tilde N}_1$.
Neglecting this contribution we rewrite Eq. (\ref{n22}) as
\begin{eqnarray*}
{\tilde N}_2 (k, Y) = - \int \frac{d \lambda_1 
\, d \lambda_2}{(2 \pi i)^2} \, C_{\lambda_1} \, C_{\lambda_2} \, 
\left(\frac{k}{\Lambda}\right)^{\lambda_1 + \lambda_2} \, \frac{1}{2 
[\chi (- \lambda_1) + \chi (- \lambda_2) - \chi (- \lambda_1 - 
\lambda_2)]} 
\end{eqnarray*}
\begin{eqnarray}\label{n23}
\times \exp \left( 2 \, \frac{\alpha N_c}{\pi} \, Y \, [\chi (- \lambda_1) 
+ \chi (- \lambda_2)] \right) .
\end{eqnarray}
Here again we can perform the integration over $\lambda_1$ and
$\lambda_2$ in the saddle point approximation. The saddle points of
$\lambda_1$ and $\lambda_2$ integrations would have been independent
if it were not for the $\chi (- \lambda_1 - \lambda_2)$ in the
denominator. The problem is that $\chi (\lambda)$ has a pole at
$\lambda = 2$, and since the saddle points of $\lambda_1$ and
$\lambda_2$ integrations are located in the vicinity of $-1$ [see Eq.
(\ref{sp})] this could influence the exact location of the saddle
points of the integral. However, as one can see analyzing the saddle
points of Eq.  (\ref{n23}), if $\ln^2 (k/\Lambda) > \alpha Y N_c$ the
effect of $\chi (- \lambda_1 - \lambda_2)$ could be safely neglected.
This imposes an additional constraint on the possible values of
transverse momentum $k$. Remember that in order to perform the saddle
point integration in Eq.  (\ref{nmel}) we had to assume that $\ln
(k/\Lambda) \ll \alpha Y N_c$.  These two conditions specify the range
of $k$ in which our analysis is applicable. The condition $\ln^2
(k/\Lambda) > \alpha Y N_c$ is not very hard to satisfy, since, after
all, the transverse momentum $k$ should be large enough for
perturbative QCD to be applicable, i.e. , $\ln (k/\Lambda) \gg 1$. We
will return to this question below.

Performing the saddle point integration in Eq. (\ref{n23}) in the
spirit of the above discussion one obtains
\begin{equation}\label{p2}
{\tilde N}_2 (k, Y) = - \frac{a (k, Y)}{1 + 8 \, a (k, Y) \,  \ln 2} 
\, P_1 (k, Y)^2,
\end{equation}
with $a (k, Y)$ defined in Eq. (\ref{a}). In arriving at Eq.
(\ref{p2}) we had to also include the non--singular terms in $a$ in
the denominator of Eq. (\ref{n23}). This was done for the reasons
which will become apparent below, when we will explore the
convergence of the series we will obtain.

We can continue construction of our perturbative solution by finding
the third order correction ${\tilde N}_3 (k, Y)$. Writing ${\tilde N}
(k, Y) = {\tilde N}_1 (k, Y) + {\tilde N}_2 (k, Y) + {\tilde N}_3 (k,
Y)$ we obtain the following equation for ${\tilde N}_3 (k, Y)$ from Eq.
(\ref{bsp})
\begin{equation}\label{n3eq}
\frac{\partial {\tilde N}_3 (k, Y)}{ \partial Y} = \frac{2 \alpha
    N_c}{\pi} \, \chi \left( - \frac{\partial}{\partial \ln k} \right)
  \, {\tilde N}_3 (k, Y) - \frac{\alpha N_c}{\pi} \, 2 \, {\tilde N}_1 (k,
  Y) \, {\tilde N}_2 (k, Y) .
\end{equation}
The solution to Eq. (\ref{n3eq}) is given by 
\begin{equation}\label{n31}
{\tilde N}_3 (k, Y) = - \frac{\alpha N_c}{\pi} \int_0^Y dy \exp \left[ 
\frac{2 \alpha N_c}{\pi} \, (Y - y) \,\chi \left( - \frac{\partial}
{\partial \ln k} \right)\right] [ \, 2 \, {\tilde N}_1 (k, y) \, {\tilde N}_2 
(k, y) ] ,
\end{equation}
similarly to Eq. (\ref{n21}). Substituting ${\tilde N}_1 (k, y)$ from
Eq. (\ref{nmel}) and ${\tilde N}_2 (k, y)$ from Eq. (\ref{n23}),
performing the integration over $y$ and neglecting the two pomeron
contribution which arises in the same way as one pomeron contribution
appeared in Eq. (\ref{n22}) we reduce Eq. (\ref{n31}) to the following
form
\begin{eqnarray*}
{\tilde N}_3 (k, Y) =  2 \, \int \frac{d \lambda_1 
\, d \lambda_2 \, d \lambda_3}{(2 \pi i)^3} \, C_{\lambda_1} \, 
C_{\lambda_2} \, C_{\lambda_3} \, \left(\frac{k}{\Lambda}\right)^{\lambda_1 
+ \lambda_2 + \lambda_3} \, \frac{1}{2 [\chi (- \lambda_2) + \chi (- \lambda_3) 
- \chi (- \lambda_2 - \lambda_3)]}  
\end{eqnarray*}
\begin{eqnarray}\label{n33}
\times  \frac{1}{2 [\chi (- \lambda_1) + \chi (- \lambda_2) + \chi (- \lambda_3) 
- \chi (- \lambda_1 - \lambda_2 - \lambda_3)]} \, \exp \left( 2 \, 
\frac{\alpha N_c}{\pi} \, Y \, [\chi (- \lambda_1) 
+ \chi (- \lambda_2) + \chi (- \lambda_3)] \right) .
\end{eqnarray}
After saddle point integrations Eq. (\ref{n33}) yields
\begin{eqnarray}
{\tilde N}_3 (k, Y) = \frac{a (k, Y)}{2 \, (2 \, \ln 2 + 1) \, (1 + 8 \, 
a (k, Y) \, \ln 2)} \, P_1  (k, Y)^3.
\end{eqnarray}

Now we are ready to write down the general expression for the
perturbation series of (\ref{ps}). An obvious generalization of Eqs.
(\ref{n21}) and (\ref{n31}) allows us to write
\begin{equation}\label{neq}
{\tilde N}_n (k, Y) = - \frac{\alpha N_c}{\pi} \int_0^Y dy \exp \left[ 
\frac{2 \alpha N_c}{\pi} \, (Y - y) \,\chi \left( - \frac{\partial}
{\partial \ln k} \right)\right] \, \left( \, \sum_{m=1}^{n-1} \, {\tilde N}_m (k, y)
\, {\tilde N}_{n-m} (k, y) \right).
\end{equation}
As one could see from above the term of order $n$ in the series is
proportional to $P_1^n$. Each pomeron splitting through the
integration over $y$ introduces a denominator of the type present in
Eqs.  (\ref{n23}) and (\ref{n33}).  The series becomes
\begin{equation}\label{psol}
{\tilde N} (k, Y) = \sum_{n=1}^\infty \, f_n \, P_1 (k,Y)^n,
\end{equation}
where the coefficients of the series should be determined from the
relationship
\begin{equation}\label{fn}
f_n = a_n \sum_{m=1}^{n-1} \, f_m \, f_{n-m}, \hspace*{1cm} f_1 =1,
\end{equation}
with
\begin{equation}\label{an}
a_n = \left\{ \begin{array}{c} - \frac{1}{2} \, n \, a (k,Y) 
\{ 1 + n \, a (k,Y) \, [n \, 2 \, \ln 2 - \psi (1) + \psi (n/2) ] \}^{-1}, 
\hspace*{1cm} \mbox{even} \, \,  n, \\ \\ \\  - \frac{1}{2} \, \{ n \, 2 \, \ln 2 
- \psi (1) + \psi (n/2) \}^{-1}, \hspace*{1cm} \mbox{odd} \, \, n .
\end{array} \right.
\end{equation}

Eqs. (\ref{psol}), (\ref{fn}) and (\ref{an}) provide us with the exact
solution of Eq. (\ref{bsp}) in the kinematic region where the series
of Eq.  (\ref{psol}) is convergent. Since each term ${\tilde N}_n (k,
Y)$ is given by Eq. (\ref{neq}) the infinite sum of these terms
(\ref{psol}) satisfies Eq. (\ref{bsp}) giving the exact solution to it
in the region where it is finite. To obtain Eq. (\ref{an}) we had to
assume that $n \, a (k, Y) \ll 1$ (as $a (k,Y) \ll 1$) and expand the
$\chi $ functions in the denominators to the first non-singular terms
in $a (k,Y)$. Eq. (\ref{psol}) represents an expansion of the
amplitude in terms of multiple hard pomeron exchanges, the $n$th term
in the series corresponding to the $n$-pomeron exchange contribution.

\begin{figure}
\begin{center}
  \epsfxsize=15cm
\leavevmode
\hbox{ \epsffile{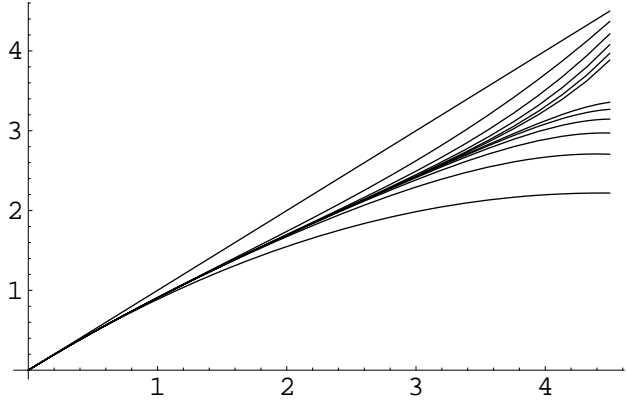}}
\end{center}
\caption{Plot of partial sums ($S_n$) of the series of Eq. (\ref{psol}) as 
  functions of $P_1$, keeping $a=0.3$ for $n=1, \ldots ,12$. The thick
  line formed by the partial sums corresponds to the true value of
  ${\tilde N} (k,Y)$. The straight line represents the single BFKL
  pomeron exchange contribution. }
\label{uni}
\end{figure}

Eqs. (\ref{fn}) and (\ref{an}) yield us with the prescription of
calculation of an arbitrary term in the series for ${\tilde N} (k,Y)$.
One could explore the effects of the higher order corrections to the
BFKL pomeron arising from Eq. (\ref{bsp}) by plotting the partial sums
of the series of Eq.  (\ref{psol}). Defining $S_n = \sum_{m=1}^n \,
f_m \, P_1^m$, we plot $S_n$ as a function of $P_1$ in Fig. \ref{uni}
for $n=1, \ldots , 12$.  We fixed $a$ to be a constant equal to $0.3$
for simplicity. The single BFKL pomeron contribution corresponds to
the first term in the series, $S_1 = P_1$, and is given by the
straight line in Fig. \ref{uni}, because we plotted $P_1$ as an
argument on the horizontal axis. Each partial sum $S_n$ gives us some
approximation to the exact result for ${\tilde N}$, and then starting
from some value of $P_1$ starts diverging. Inclusion of higher orders
gives better approximations. That way the thick line formed by the
overlap of the curves corresponding to different partial sums
represents the true value of ${\tilde N}$, the solution of Eq.
(\ref{bsp}) to which the series is converging. One can see that the
exact solution shown by that thick line lies below the single BFKL
pomeron contribution (straight line), deviating from it more and more
as the energy (and, therefore, $P_1$) increases. Thus the multiple
pomeron contributions start unitarizing the BFKL pomeron. An analogous
conclusion was reached by Salam in \cite{Salam} through numerical
study of the multiple pomeron exchanges in the onium--onium scattering
in the framework of the dipole model \cite{Mueller3}.

To find the radius of convergence and the high energy asymptotic
behavior of the series in Eq. (\ref{psol}) let us study the terms
corresponding to large values of $n$. These terms describe the
multiple pomeron exchange contributions. We would like to explore very
large $n$, such that $n^2 a (k,Y) \gg 1$. Then the coefficients $a_n$
determining the recursion relations in (\ref{fn}) for the coefficients
of the series become
\begin{equation}
a_n = - \frac{1}{n \, 4 \, \ln 2}.
\end{equation}

Thus, redefining
\begin{equation}\label{resc}
f_n = \left( \frac{- 1}{4 \, \ln 2} \right)^{n-1} \tilde{f}_n,
\end{equation}
we obtain the following recursion relations instead of Eq. (\ref{fn})
\begin{equation}\label{recur}
\tilde{f}_n = \frac{1}{n} \sum_{m=1}^{n-1} \, \tilde{f}_m \, \tilde{f}_{n-m}, 
\hspace*{1cm} \mbox{large} \, n.
\end{equation}
For very large $n$ Eq. (\ref{recur}) could be satisfied up to order
$1/n$ corrections with the ansatz 
\begin{equation}\label{ans}
\tilde{f}_n = \left( \frac{1}{r} \right)^n + o \left( \frac{1}{n} \right) , 
\end{equation}
where $r$ could be some arbitrary constant. Here we have to point out
that since Eq. (\ref{recur}) is written for very large $n$ we can not
use $f_1 = 1$ as a ``normalization'' condition anymore and that way we
have lost the information which would have allowed us to fix the value
of $r$.  In principle $r$ should depend on $a (k,Y)$, but this
function is a much slower varying function of $k$ and $Y$ than $P_1
(k,Y)$, so we assume $r$ to be constant in our analysis. If we fix
$\tilde{f}_1 = 1$ then numerical estimates show that $r \approx 1.4$,
which allows one to hope that for asymptotics of Eq. (\ref{fn}) $r$ is
not very large and does not introduce a large correction to our
analysis.

Using the ansatz of Eq. (\ref{ans}) together with Eq. (\ref{resc}) in
Eq. (\ref{psol}) we conclude that the constructed perturbation series
is convergent as long as
\begin{equation}\label{conv}
\frac{P_1 (k,Y)}{r \, 4 \, \ln 2} < 1.
\end{equation}
Outside the region specified by the condition of Eq. (\ref{conv}) the
series in Eq. (\ref{psol}) is not convergent anymore. Multiple pomeron
exchange contributions corresponding to large values of $n$ become
much larger and, consequently, more important than the one- or
two-pomeron exchanges. Defining the saturation momentum scale $Q_s$ by
the following condition
\begin{equation}
\frac{P_1 (Q_s ,Y)}{r \, 4 \, \ln 2} = 1
\end{equation}
we obtain (for not very large $Q_s$) \cite{Mueller6}
\begin{equation}
Q_s (Y) = \Lambda \, \frac{C_{-1}}{r \, 4 \, \ln 2} \, \frac{\exp [ (\alpha_P 
- 1)Y] }{\sqrt{14 \alpha N_c \zeta (3) Y}}. 
\end{equation}
Using this quantity we can say that for the transverse momenta of the
quark--antiquark pair greater than the saturation momentum, $k > Q_s$,
the series converges. For the transverse momenta below the saturation
scale, $k < Q_s$, that is, inside the saturation region, the series
diverges ceasing to be a reliable approximation to the solution of Eq.
(\ref{bsp}).

One might wonder whether the condition $\ln^2 (k/\Lambda) > \alpha N_c
Y$, which we needed to perform the saddle point integrations above,
would prevent us from exploring the high energy asymptotics with the
series of Eq. (\ref{psol}) by putting an upper limit on the possible
values of $Y$ for fixed $k$. Really, this condition provides us with
the limit on the range of rapidities for the perturbative solution we
found
\begin{equation}\label{ysp}
Y < Y_{sp} \sim \frac{1}{\alpha N_c} \, \ln^2 \frac{k}{\Lambda}. 
\end{equation}
However we also have to note that the series (\ref{psol}) is
convergent as long as $k > Q_s (Y) \sim \alpha^2 \Lambda e^{(\alpha_P
  - 1) Y}$, i.e., outside of the saturation region. This yields us
with another constraint on $Y$
\begin{equation}\label{ysat}
Y < Y_{sat} \sim \frac{1}{\alpha_P - 1} \, \ln \frac{k}{\alpha^2 
\Lambda}. 
\end{equation}
Now one can see that the two conditions (\ref{ysp}) and (\ref{ysat})
are roughly of the same order of magnitude. Thus the condition $\ln^2
(k/\Lambda) > \alpha N_c Y$ does not impose any additional constraint
on the kinematic region of the applicability of the series in Eq.
(\ref{psol}) since the series diverges in the region where the
condition is violated anyway.

Now we want to find some approximation to the solution of Eq.
(\ref{bsp}) inside the saturation region. The saturation region is
specified by the condition that $k < Q_s (Y)$ and could be achieved
either by decreasing the transverse momentum or by increasing the
center of mass energy of the system.

First we note that one may find a correction to Eq. (\ref{ans}).
Rewriting it as
\begin{equation}\label{cor1}
\tilde{f}_n = \tilde{f}_n^{(0)} + \epsilon_n ,
\end{equation}
where $\tilde{f}_n^{(0)} = (1/r)^n$ for large $n$ and
$\tilde{f}_n^{(0)} \sim a^{n/2} \ll 1$ for small $n$, i.e., it
satisfies Eq.  (\ref{fn}) with the large $n$ asymptotics given by
leading in $n$ behavior of Eq. (\ref{ans}). Substituting Eq.
(\ref{cor1}) into Eq. (\ref{ans}), expanding it to the lowest order in
$\epsilon_n$ and neglecting $\tilde{f}_n^{(0)}$ for small values of
$n$ one obtains
\begin{equation}\label{cor2}
\epsilon_n \sim \frac{1}{n} \, \left( \frac{1}{r} \right)^n .
\end{equation}
Using Eqs. (\ref{cor2}), (\ref{ans}) and (\ref{resc}) in (\ref{psol})
we could sum up leading in $n$ terms of the series (\ref{psol})
obtaining some expressions which could be analytically continued
outside the region of convergence of the series of Eq.  (\ref{psol}).
This way we might obtain an expression for the solution of Eq.
(\ref{bsp}) inside the saturation region, that is for large $P_1$.
While the leading in $n$ part of the series given by Eq.  (\ref{ans})
gives a function which goes to a constant for large $P_1$, the
subleading terms given by Eq. (\ref{cor2}) yield us with
\begin{equation}\label{sans}
\sum_{n=1}^\infty \frac{(-1)^{n-1}}{n} \, \left( \frac{P_1}{r \, 4 \, 
\ln 2} \right)^n = \ln \left( 1 + \frac{P_1}{r \, 4 \, \ln 2} \right).
\end{equation}
Analytically continuing $\ln (1 + \frac{P_1}{r \, 4 \, \ln 2})$ in the
region of large $P_1$ we obtain for asymptotically large energies $\ln
\frac{P_1}{r \, 4 \, \ln 2}$. Of course we do not know whether the
higher order corrections in $1/n$ would not give us some other
function of $P_1$ which would be large for large $P_1$ possibly
interfering with our $\ln P_1$ ansatz.  Therefore $\ln P_1$ should be
considered only as a a possible high energy asymptotics of $\tilde{N}$
and the above argument should not be taken as a proof of this
statement.

To justify that the above ansatz of $\tilde{N} = \ln \frac{P_1}{r \, 4
  \, \ln 2} = \ln \frac{Q_s}{k}$ is the solution of Eq.  (\ref{bsp})
at asymptotically high energies we first note that
\begin{equation}\label{log}
\tilde{N} (k, Y) = \ln \frac{M}{k},
\end{equation}
with $M$ some large momentum scale ($k < M$) satisfies the equation
\begin{equation}\label{hen}
2 \, \chi \left( - \frac{\partial}{\partial \ln k} \right) \, {\tilde N} 
(k, Y) - {\tilde N}^2 (k, Y) =0.
\end{equation}
Really, rewriting
\begin{equation}
\ln \frac{M}{k} = -  \int_C \frac{d \lambda}{2 \pi i} \, \frac{1}{\lambda^2} 
\left( \frac{k}{M} \right)^\lambda ,
\end{equation}
where the contour $C$ is a circle of radius less than one centered
around $\lambda = 0$, one can see that since $\chi (- \lambda)$ also
has a pole at $\lambda = 0$ then
\begin{equation}
2 \, \chi \left( - \frac{\partial}{\partial \ln k} \right) \, \ln \frac{M}{k} 
= \ln^2 \frac{M}{k}.
\end{equation}
We can assume that in the saturation region $\tilde{N} (k,Y)$ ceases
to depend on $Y$. Then Eq. (\ref{bsp}) reduces to Eq. (\ref{hen}). To
find the most general solution of Eq. (\ref{hen}) we represent
$\tilde{N}$ as a series
\begin{equation}\label{hens}
\tilde{N} (k,Y) = \sum_{n,m=0}^\infty c_{n m} \left( \frac{k}{M} \right)^n 
\, \ln^m  \frac{M}{k}
\end{equation}
and substitute it into Eq. (\ref{hen}) in order to find the unknown
coefficients $c_{nm}$. Using the technique outlined above for $\ln
\frac{M}{k}$ we can show that
\begin{equation}\label{chiac}
2 \, \chi \left( - \frac{\partial}{\partial \ln k} \right) \, \left[  
\left( \frac{k}{M} \right)^n  \, \ln^m  \frac{M}{k} \right] = 
\left\{ \begin{array}{c}  \frac{2}{m + 1}  \left( \frac{k}{M} \right)^n  
\, \ln^{m+1}  \frac{M}{k}, \hspace*{1cm} \mbox{even} \, \,  n, \\ \\ 
2 \, \chi (-n) \left( \frac{k}{M} \right)^n  \, \ln^m  \frac{M}{k} , 
\hspace*{1cm} \mbox{odd} \, \, n.
\end{array} \right.
\end{equation}
With the help of Eq. (\ref{chiac}) one can show that in order for the
series of Eq. (\ref{hens}) to satisfy Eq. (\ref{hen}) all of its
coefficients must be zero with the exception of $c_{01} = 1$. Thus we
proved that the most general energy independent solution of Eq.
(\ref{bsp}) is given by Eq. (\ref{log}).

At asymptotically high energies there is only one large momentum scale
in the scattering of the quark--antiquark pair on a hadron or nucleus
--- the saturation scale $Q_s$. This scale is also suggested by the
ansatz of Eq. (\ref{sans}).  Therefore we should put $M = Q_s$ and use
Eq. (\ref{log}) as a solution of Eq. (\ref{bsp}) in the region when $k
< Q_s$. The derivative of this new solution with respect to rapidity
is of the order of $\partial \ln (Q_s / k)/ \partial Y \sim \alpha
N_c$ and could be neglected compared to any of the terms on the right
hand side of Eq. (\ref{bsp}), each of them being of the order of
$\alpha N_c (\alpha N_c Y)^2 $. This could be done since $\alpha N_c Y
\gg 1$.  Therefore we conclude that in the saturation region
\begin{equation}\label{asym}
\tilde{N} (k, Y) \approx \ln \frac{Q_s}{k}, \hspace*{1cm} Y \ge 
\frac{1}{\alpha_P - 1} \, \ln \, \frac{1}{\alpha^2}.
\end{equation}

Now that we know the solution for $\tilde{N} (k, Y)$ outside of the
saturation region given by Eqs. (\ref{psol}), (\ref{fn}) and
(\ref{an}) and the high energy asymptotics inside the saturation
region given by Eq. (\ref{asym}), one might want to construct the
solution in coordinate space ($N (x_\perp, Y)$) using the Fourier
transformation of Eq.  (\ref{trfm}). To do that one would have to
integrate over all values of the transverse momentum $k$ from $0$ to
$\infty$. If $x_\perp < 1 / Q_s$ then the transverse momentum, which
in Eq. (\ref{trfm}) is effectively cut off by the Bessel function and
, therefore, varies approximately from $0$ to $1 / x_\perp > Q_s$,
will go through a range of values both above and below $Q_s$. For the
first case one would have to use the perturbation series which we
constructed as $\tilde{N}$, while for the second case one might use
the asymptotic value of $\tilde{N}$ given by Eq. (\ref{asym}) ,
although a perturbative expansion around it would be more accurate.
The result would be a complicated combination of special function and
we are not going to list it here.

If $x_\perp > 1 / Q_s$ then the transverse momentum in Eq.
(\ref{trfm}) simply gets cut off by $1 / x_\perp < Q_s$ and always
stays within the saturation region. Thus one could just perform a
Fourier transformation of formula (\ref{asym}) using Eq.
(\ref{trfm}). The result is
\begin{equation}\label{nsa}
N (x_\perp , Y) \approx 1 , \hspace*{1cm} Y \ge 
\frac{1}{\alpha_P - 1} \, \ln \, \frac{1}{\alpha^2}.
\end{equation}
This corresponds to the blackness of the total cross section of the
quark--antiquark pair on the target nucleus, since it is given by
\begin{equation}\label{geo}
\sigma_{q \overline{q} A}^{tot} = 2 \, \int d^2 b_0 N (x_\perp, b_0, Y)
 \approx 2 \pi R^2
\end{equation}
for $N=1$ and for a cylindrical nucleus of radius $R$ as described
above. That way the total cross section is independent of energy at
asymptotically high energies and is completely unitary. It reaches its
``geometrical'' limit (\ref{geo}) which could be predicted even from
quantum mechanics.

\begin{figure}
\begin{center}
\epsfxsize=15cm
\leavevmode
\hbox{ \epsffile{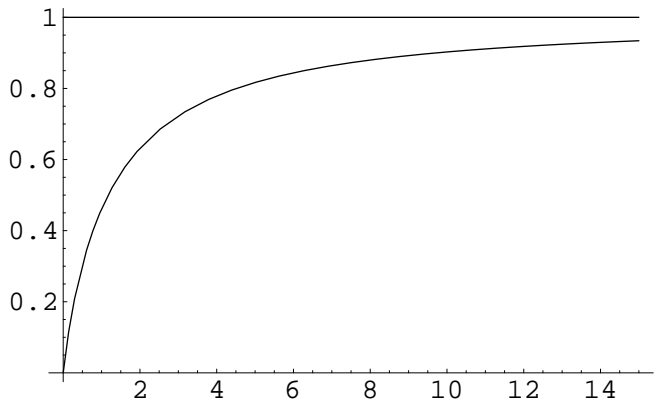}}
\end{center}
\caption{Plot of the qualitative behavior of the forward $q \overline{q}$
  pair scattering amplitude on a nucleus, $N (x_\perp , Y)$ , as a
  function of $x_\perp Q_s (Y)$ (one BFKL pomeron exchange
  contribution). At small values of $x_\perp Q_s$ the structure
  function is linear in $x_\perp Q_s$, reproducing BFKL pomeron. As
  energy increases the forward amplitude $N$ saturates to a constant.}
\label{nsat}
\end{figure}

That way we have shown that $N (x_\perp , Y)$ given by the solution of
Eq. (\ref{bsp}) behaves like a single BFKL pomeron exchange
contribution at moderately high energies ($Y \sim 1/\alpha$), which
follows from the perturbation series we have constructed, and as
energy increases to very high quantities ($Y \ge \frac{1}{\alpha_P -
  1} \, \ln \, \frac{1}{\alpha^2}$), saturates to a constant
independent of energy. A qualitative sketch of $N (x_\perp , Y)$ as a
function of $x_\perp Q_s (Y)$ (one pomeron contribution) is shown in
Fig. \ref{nsat}.

\section{Asymptotic behavior of the $F_2$ structure function}

For a large cylindrical nucleus the $F_2$ structure function is given by 
\begin{equation}\label{f21}
F_2 (x, Q^2) = \frac{Q^2 R^2}{8 \pi^2 \alpha_{EM}} \int d^2 x_{01} \, dz \, 
\Phi (x_{01}, z) \, N (x_{01}, Y),
\end{equation}
which follows from Eq. (\ref{f2}). We can rewrite Eq. (\ref{f21}) in
momentum space using Eq. (\ref{trfm})
\begin{equation}\label{f22}
F_2 (x, Q^2) = \frac{Q^2 R^2}{4 \pi \alpha_{EM}} \int d^2 k \, dz \, 
\tilde{\Phi} (k, z) \, \tilde{N} (k, Y),
\end{equation} 
where
\begin{equation}
\tilde{\Phi} (k, z) = \int \frac{d^2 x_{01}}{(2 \pi)^2} \, e^{i {\bf k} 
\cdot {\bf x}_{01}} \, x^2_{01} \,  \Phi (x_{01}, z).
\end{equation} 
Employing Eqs. (\ref{ft}) and (\ref{fl}) we obtain
\begin{eqnarray*}
\tilde{\Phi} (k, z) = \frac{\alpha_{EM} N_c}{\pi^2} \left\{ [z^2 + (1-z)^2] \,
\frac{2}{3 a^2} \,  _2F_1 \left( 2, 3, 2.5, - \frac{k^2}{4 a^2} \right) + \right.
\end{eqnarray*}
\begin{eqnarray}\label{hyp}
\left. + 4 \, Q^2 \, z^2 (1-z)^2 \, \frac{1}{15 a^6} \, \left[ 5 \, a^2 \, 
_2F_1 \left( 2, 2, 2.5, - \frac{k^2}{4 a^2} \right) - 2 \, k^2 \,  _2F_1 \left( 
3, 3, 3.5, - \frac{k^2}{4 a^2} \right) \right] \right\},
\end{eqnarray}
where $_2F_1$ is a hypergeometric function. 

Eq. (\ref{f22}) shows that in order to write down an expression for
the $F_2$ structure function one has to know $\tilde{N} (k,Y)$ in the
areas above and below saturation. The situation is similar to the
Fourier transformation of the previous section. The wave function
$\tilde{\Phi} (k,z)$ given by Eq. (\ref{hyp}) becomes very small for
transverse momenta $k > a = Q \sqrt{z (1-z)}$, effectively providing
an upper cutoff on the $k$ integration. Thus if $Q > Q_s$ the integral
over $k$ includes $k < Q_s $ and $k > Q_s$ and to calculate it we have
to use the solution for $\tilde{N}$ both inside and outside of the
saturation region. We can again propose the approximation outlined in
Sect. II for calculation of $F_2$, which consists of using
perturbation series of Eq. (\ref{psol}) outside saturation region and
approximating the solution inside that region by Eq. (\ref{asym}). The
result would also include a complicated list of special functions
which we are not going to list here.  We have to admit that numerical
solution of Eq. (\ref{eqN}) could give a more precise results for
$F_2$.  Here we can just point out that at moderately high energies
the behavior of the $F_2$ structure function will be dominated by
single BFKL pomeron exchange, providing us with the following formula
which could be obtained by substituting the first term in the series
of Eq. (\ref{psol}) into Eq. (\ref{f22})
\begin{equation}\label{f2b}
F_2 (x, Q^2) = \frac{11}{256} \, S_\perp N_c Q \Lambda \, \frac{C_{-1}}
{4 \, \ln 2} \, \frac{\exp [ (\alpha_P - 1)Y] }{\sqrt{14 \alpha N_c 
\zeta (3) Y}}, \hspace*{1cm} Y \sim \frac{1}{\alpha},
\end{equation}
where we have neglected the transverse momentum diffusion term in the
exponent of Eq. (\ref{n1}) and $S_\perp = \pi R^2$. Eq. (\ref{f2b}) is
valid only for moderately high energies, when the saturation momentum
$Q_s$ is small and we can neglect the $k$ integration in Eq.
(\ref{f22}) below that scale.

When $Q < Q_s$ or the energy gets very high to sufficiently increase
$Q_s$ the integral over $k$ in Eq. (\ref{f22}) is limited to $k < Q <
Q_s$, so that we can use Eq.  (\ref{asym}) for $\tilde{N} (k,Y)$. The
result yields
\begin{equation}
F_2 (x, Q^2) = \frac{Q^2 R^2 N_c}{3 \pi^2} \, \left( \ln \frac{Q_s}{Q} + 
\frac{13}{12} \right), \hspace*{1cm} Q < Q_s.
\end{equation}
As energy becomes very large, corresponding to $Y \sim
\frac{1}{\alpha_P - 1} \ln \frac{1}{\alpha^2} $, $\ln \frac{Q_s}{Q}$
becomes approximately equal to $(\alpha_P - 1) Y$. Therefore 
\begin{equation}\label{f2a}
F_2 (x, Q^2) \approx \frac{Q^2 R^2 N_c}{3 \pi^2} \, (\alpha_P - 1) Y , 
\hspace*{1cm} Y \ge \frac{1}{\alpha_P - 1} \ln \frac{1}{\alpha^2}.
\end{equation}
We conclude that $F_2 \sim \ln s$ at asymptotically high energies.
This conclusion may seem to be a little unusual, so we will reproduce
the same result using more conventional coordinate space language.  As
was shown in the previous section the forward amplitude of the $q
\overline{q}$ pair scattering on a nucleus is $N (x_\perp, Y) = 1$,
when $x_\perp > 1/Q_s$. When $x_\perp < 1/Q_s$ one could argue that
the amplitude is roughly proportional to some positive power of
$x_\perp$, at least $N (x_\perp, Y) \sim x_\perp$, as given by one
pomeron exchange. Thus, for $Q < Q_s$ we can neglect the part of the
integral with $x_{01} < 1/Q_s$ in Eq. (\ref{f21}), since the wave
function $\Phi$ in Eq.  (\ref{f21}) behaves like $\frac{1}{x_{01}^2}$
and is not singular enough to make that portion of the integral
significant. That way, expanding the modified Bessel functions in
$\Phi$ of Eq.  (\ref{ft}) and integrating over $z$ we obtain
\begin{equation}\label{f23}
F_2 (x, Q^2) \approx \frac{Q^2 R^2 N_c}{3 \pi^2} \, \int_{1/Q_s}^{1/Q} \, 
\frac{d x_{01}}{x_{01}}. 
\end{equation}
The upper cutoff on the $x_{01}$ integration in Eq. (\ref{f23}) is
provided by the fact that the modified Bessel functions fall off
exponentially at the large values of the argument. Integrating over
$x_{01}$ in Eq. (\ref{f23}) we arrive at Eq. (\ref{f2a}).

That way we have shown that the $F_2$ structure function given by the
solution of Eq. (\ref{neq}) is proportional to the single BFKL pomeron
exchange contribution (\ref{f2b}) at moderately high energies with $Y
\sim 1/\alpha$, and, as energy increases the $F_2$ structure function
unitarizes, becoming linearly proportional to $\ln s$ (\ref{f2a}). We
stress again that even though the cross section of the
quark--antiquark pair of a fixed transverse size $x_\perp$ scattering
on a nucleus saturates to a constant at large $Y$, the integral over
$x_\perp$ makes $F_2$ depend on $Y$, such that $F_2 \sim Y$ at very
large $Y$.

\section{Application to diffraction}

\begin{figure}
\begin{center}
\epsfxsize=12cm
\leavevmode
\hbox{ \epsffile{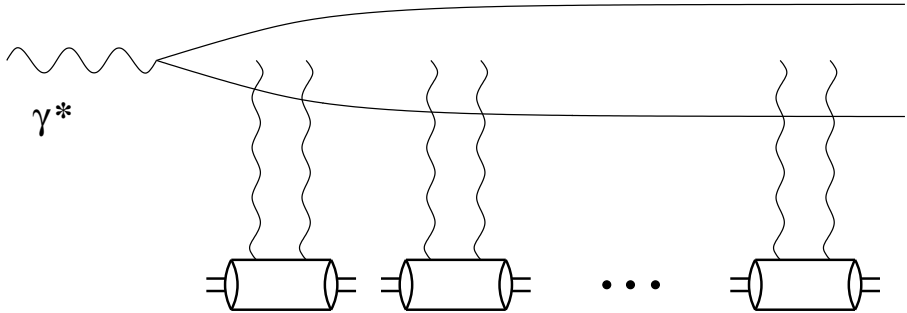}}
\end{center}
\caption{Quasi--elastic deep inelastic scattering as pictured in [25].}
\label{ampl}
\end{figure}

In \cite{mL} a diffractive deep inelastic scattering process was
considered, an amplitude of which is shown in Fig. \ref{ampl}. A
virtual photon scatters on a nucleus elastically, leaving it intact.
Two quarks in the final state in Fig. \ref{ampl} may become a pair of
jets or form a vector meson. These type of processes were called
quasi--elastic virtual photoproduction in \cite{mL}. 

The interactions between the quark--antiquark pair and the nucleus
were taken in the quasi--classical approximation, similar to
\cite{qc}.  Each nucleon in \cite{mL} would interact with the $q
\overline{q}$ pair by a two gluon exchange. Strictly speaking this
approximation corresponds to not very high energies, when the
logarithms of $s$ resulting from QCD evolution are not large enough to
become important.  That is why the approximation is called
quasi--classical. In \cite{mL} it provided us with a simple Glauber
type expression for the diffractive, or, more correctly,
quasi--elastic structure function $F_2^D$ [cf. Eq. (20) in \cite{mL}]
\begin{eqnarray*}
  F_2^D (x, Q^2) = \frac{Q^2 N_c}{\pi (2 \pi)^3} \int d^2 b_0 \, \,
  d^2 x_{01} \, \int_0^1 dz \left\{ a^2 K_1^2 (x_{01} a) [z^2 + (1 -
    z)^2] + 4 Q^2 z^2 (1 - z)^2 K_0^2 (x_{01} a) \right\}
\end{eqnarray*}
\begin{eqnarray}\label{f2diff0}
  \times \gamma^2 ({\bf x}_{01},{\bf b}_0),
\end{eqnarray}
where we again assumed that the quarks are massless and have only one
flavor. $\gamma ({\bf x}_{01},{\bf b}_0)$ is the Glauber propagator of
the quark--antiquark pair, which for the case of a cylindrical nucleus
is given by Eq.  (\ref{gla}). One can obtain Eq. (\ref{f2diff0}) from
Eq. (\ref{f2}) by first noting that the quasi--classical limit
corresponds to putting $Y=0$ (no evolution in rapidity) in it and $N
({\bf x}_{01},{\bf b}_0, Y = 0) = \gamma ({\bf x}_{01},{\bf b}_0)$.
Recalling that, as was argued in \cite{mL}, one can rewrite the total
and elastic cross sections for DIS on a nucleus in terms of the
$S$-matrix at a given impact parameter of the collision, $S(b)$,
yielding
\begin{mathletters}
\begin{eqnarray}\label{stot}
\sigma_{tot} = 2 \int d^2 b \, \, [1 - S(b)] 
\end{eqnarray}
\begin{eqnarray}\label{sel}
\sigma_{el} =  \int d^2 b \, \, [1 - S(b)]^2 
\end{eqnarray}
\end{mathletters}
and associating $F_2$ with $\sigma_{tot}$ and $F_2^D$ with
$\sigma_{el}$, we can easily derive Eq. (\ref{f2diff0}) from Eq.
(\ref{f2}) \cite{mL}.

\begin{figure}
\begin{center}
\epsfxsize=10cm
\leavevmode
\hbox{ \epsffile{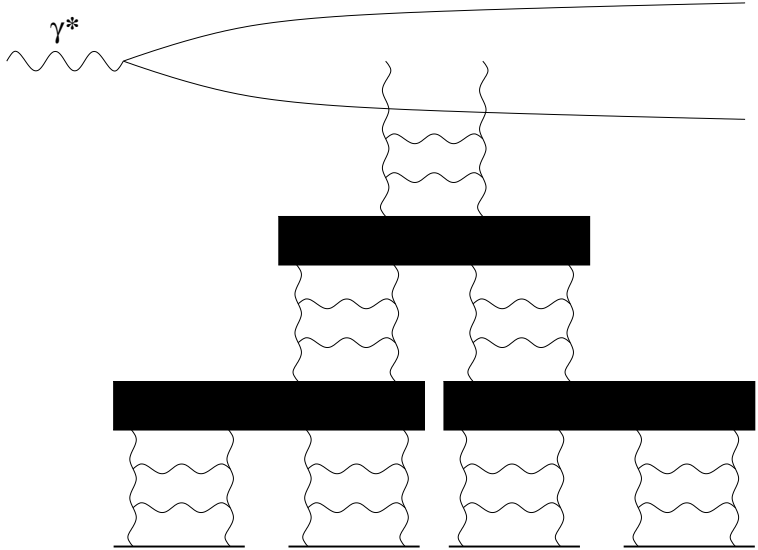}}
\end{center}
\caption{Quasi--elastic scattering including multiple pomeron exchanges, 
as envisioned in our approach.}
\label{diffr}
\end{figure}

Now let us imagine our quasi--elastic process at very high energies.
Then, strictly speaking, one has to include the evolution in rapidity
in the structure function $F_2^D$. We still do not want to have
particles being produced in the central rapidity region. One could
model the interaction by a single pomeron exchange between the
quark--antiquark pair and the nucleus, similarly to what was done
previously in \cite{g2} and references mentioned there for elastic
scattering. Here we are going to propose inclusion of multiple pomeron
exchanges in $F_2^D$. Since the above formalism allows us to sum up
pomeron fan diagrams in the leading logarithmic approximation, we can
just insert this evolution in the amplitude of the quasi--elastic
process. The diagram of the process is depicted in Fig. \ref{diffr}.
The diffractive rapidity gap between the nucleus in the final state
and the remains of the $q \overline{q}$ pair results not only from a
single pomeron exchange, but also from multiple pomeron exchanges.
Still nothing is produced in the central rapidity region.

The corresponding expression for the diffractive structure function
$F_2^D$ is easy to derive. We know the expression for the total $F_2$
structure function in Eq. (\ref{f2}), which includes all the discussed
multiple pomeron exchanges.  One can easily argue along the lines
outlined in deriving Eq. (\ref{f2diff0}), that for the case of
multiple pomeron exchanges we again have a relationship between the
total and elastic cross sections similar to the one given by Eqs.
(\ref{stot}) and (\ref{sel}). That way we conclude that to obtain a
formula for $F_2^D$ one has to substitute $\gamma ({\bf x}_{01},{\bf
  b}_0)$ in Eq. (\ref{f2diff0}) with $N ({\bf x}_{01},{\bf b}_0 , Y)$.
The result is
\begin{eqnarray*}
  F_2^D (x, Q^2) = \frac{Q^2 N_c}{\pi (2 \pi)^3} \int d^2 b_0 \, \,
  d^2 x_{01} \, \int_0^1 dz \left\{ a^2 K_1^2 (x_{01} a) [z^2 + (1 -
    z)^2] + 4 Q^2 z^2 (1 - z)^2 K_0^2 (x_{01} a) \right\}
\end{eqnarray*}
\begin{eqnarray}\label{f2diff}
  \times \, N^2 ({\bf x}_{01},{\bf b}_0 , Y),
\end{eqnarray}
where $N ({\bf x}_{01},{\bf b}_0 , Y)$ is, in the large $N_c$ limit, a
solution of Eq.  (\ref{eqN}). Eq. (\ref{f2diff}) is our master formula
for the diffractive (quasi--elastic) structure function. When $Y=0$
(no evolution) it reduces to Eq. (\ref{f2diff0}).

Using the results of Sect. II we can derive the asymptotics of the
diffractive structure function. At moderately high energies the one
pomeron exchange contribution dominates the diffractive amplitude of
Fig. \ref{diffr}. Again, similar to the previous section we assume
that since at $Y \sim 1/\alpha$ the saturation scale $Q_s$ is small
one can neglect the $k$ integration below $Q_s$. Performing a Fourier
transformation (\ref{trfm}) of the first term in the series of Eq.
(\ref{psol}) and substituting the result into Eq. (\ref{f2diff}) we
obtain for a cylindrical nucleus
\begin{equation}\label{f2db}
F_2^D (x, Q^2) = \frac{S_\perp N_c \Lambda^2}{3 \pi^3} \, \frac{C_{-1}^2}
{(4 \, \ln 2)^2} \, \frac{\exp [ 2 (\alpha_P - 1) Y] }{14 \alpha N_c 
\zeta (3) Y} \, Y, \hspace*{1cm} Y \sim \frac{1}{\alpha},
\end{equation}
where we had to cut off the $z$ integral in Eq. (\ref{f2diff}) by
$e^{-Y}$ and we again neglected the transverse momentum diffusion term
in Eq. (\ref{n1}). The diffractive structure function of Eq.
(\ref{f2db}) behaves like a square of a single pomeron exchange
contribution, similar to what was found for elastic parton scattering
in \cite{g2}. Even though $F_2^D \sim Q_s^2 (Y)$ in Eq. (\ref{f2db})
is growing with energy faster than $F_2 \sim Q_s (Y)$ in Eq.
(\ref{f2b}), unitarity is not violated, since both expressions are
valid in the region of relatively small $Q_s$ corresponding to $Y \sim
1/\alpha$. Thus at moderate energies $F_2^D \ll F_2$, as $Q_s \ll
\Lambda$.

We also note that formula (\ref{f2db}) was derived assuming that the
photon's virtuality $Q^2$ is not very large. Therefore, the fact that
$F_2^D$ in Eq. (\ref{f2db}) has no powers of $Q$ in the numerator,
whereas the corresponding $F_2$ given by Eq. (\ref{f2b}) is
proportional to $Q$, does {\it not} imply that $F_2^D$ is not a
leading twist effect, violating the theoretical predictions
\cite{mL,kope} and experimental observations \cite{zeus}. For the case
of large $Q$ one should perform a different saddle point approximation
for the terms in the perturbative series given by Eqs. (\ref{nmel}),
(\ref{n22}), etc., [see \cite{rys}], getting a different series
solving Eq. (\ref{bsp}) outside of the saturation region.  This should
result in concluding that $F_2^D$ given by the first term in that
series is not suppressed by any powers of $Q$ compared to $F_2$.
However, doing that is beyond the scope of this paper.

At very high energies, when $Q_s (Y)$ is large enough so that $Q < Q_s
(Y)$, the quasi--elastic structure function $F_2^D$ saturates.
Similarly to Sect. III we can derive the high energy asymptotic
behavior of Eq. (\ref{f2diff}). Putting $N (x_{01}, Y) = 1$ for
$x_{01} > 1/Q_s$ and imposing the cutoffs on the $x_{01}$ integration
in Eq. (\ref{f2diff}) analogously to what was done in deriving Eq.
(\ref{f23}) we deduce
\begin{equation}\label{f2da}
F_2^D (x, Q^2) \approx \frac{Q^2 R^2 N_c}{6 \pi^2} \, (\alpha_P - 1) Y , 
\hspace*{1cm} Y \ge \frac{1}{\alpha_P - 1} \ln \frac{1}{\alpha^2}.
\end{equation}
That way the $F_2^D$ also becomes linearly proportional to the
logarithm of center of mass energy energy at very high energies,
similarly to the total structure function $F_2$.  Comparing Eq.
(\ref{f2da}) with the high energy behavior of the $F_2$ structure
function given by Eq. (\ref{f2a}) we conclude that $F_2^D$ is still
smaller than $F_2$ and unitarity is again not violated. One can see
that at this asymptotically high energy $F_2 = 2 F_2^D$, which
corresponds to the prediction for the black cross sections resulting
from quantum mechanics.

\section{Conclusions}

In this paper we have found a solution of the evolution equation
(\ref{eqN}) which resums all pomeron fan diagrams in the leading
logarithmic limit for large $N_c$, that was proposed in \cite{me}. The
solution consists of two parts. Outside of the saturation region we
have constructed a perturbative series in momentum space given by Eqs.
(\ref{psol}), (\ref{fn}) and (\ref{an}). We have demonstrated that the
series tends to unitarize the single BFKL pomeron exchange
contribution (see Fig.  \ref{uni}). In deriving formula (\ref{bsp}),
which led to Eqs. (\ref{psol}), (\ref{fn}) and (\ref{an}), we had to
assume that DIS cross sections are slowly varying functions of the
impact parameter, which corresponds to scattering on a large nucleus.
Only in the large nucleus case we could argue that the fan diagrams
dominate the total DIS cross sections.  In order to perform the saddle
point approximation in Eqs.  (\ref{nmel}), (\ref{n22}), (\ref{n33})
and for the series (\ref{psol}) to be convergent we had to assume that
$\ln (Q/\Lambda) \ll \alpha N_c Y$ and $Q > Q_s (Y)$, restricting the
possible values of $Q$ to a certain not very broad range.
Nevertheless the abovementioned assumptions did not change the
qualitative behavior of the obtained result and the conclusions could
be easily extended to large $Q$ by performing a different saddle point
approximation.

Inside the saturation region, for $Q < Q_s$, we found an approximate
solution given by Eq. (\ref{asym}) corresponding to the saturation of
the total scattering cross section of a quark--antiquark pair of a
given transverse size on a nucleus to a constant independent of energy
(\ref{nsa}). That way we have shown that the coordinate space forward
amplitude of the $q \overline{q}$ scattering on a nucleus or hadron
grows like a BFKL pomeron exchange contribution at moderately high
energies (\ref{n1}), and, as energy gets very large it saturates to a
constant (\ref{nsa}).  A qualitative plot of $N (x_\perp, Y)$ is
presented in Fig.  \ref{nsat}.

We have to admit that in order to construct a solution of Eq.
(\ref{eqN}) in coordinate space, $N (x_\perp, Y)$, and the
corresponding structure function $F_2$ one has to have a better
knowledge of the momentum space solution inside the saturation region.
A numerical solution of Eq. (\ref{eqN}) would probably be very helpful
in determining exact values of $N (x_\perp, Y)$ and $F_2$ at
intermediately large rapidities.

We have shown that the $F_2$ structure function given by the solution
of Eq. (\ref{eqN}) grows as a power of energy at moderately high
energies, corresponding to rapidities of the order of $Y \sim
1/\alpha$. This behavior, presented in Eq.  (\ref{f2b}), reproduces
the usual single BFKL pomeron exchange contribution to the structure
function. As energy grows very high, corresponding to the rapidities
of the order of $Y \gsim [1 / (\alpha_P - 1)] \ln (1 / \alpha^2 )$,
the $F_2$ structure functions unitarizes, becoming linearly
proportional to $\ln s$, which satisfies Froissart bound. This is
shown in Eq. (\ref{f2a}).  Thus we have shown that the BFKL pomeron on
the nucleus is unitarized by summation of pomeron fan diagrams.

We also note that we derived an expression for the quasi--elastic
(diffractive) structure function $F_2^D$ presented in Eq.
(\ref{f2diff}), which also includes all multiple pomeron exchanges. We
have shown that at moderately high energies the obtained expression
behaves like a square of the BFKL pomeron contribution [Eq.
(\ref{f2db})] and at very high energies it saturates and becomes a
linear function of $\ln s$ [Eq. (\ref{f2da})], being equal to a half
of the structure function $F_2$, as expected for black cross sections.

In \cite{me} it was argued that the dipole model provides us with the
techniques that allow one to resum pomeron fan diagrams at any
subleading logarithmic order in the large $N_c$ limit. That is, if one
calculates the next-to-leading order kernel in the dipole model, the
resulting equation for the generating functional $Z$ [see
\cite{Mueller1,Mueller2,Mueller3}] would provide us with the equation
resumming all NLO pomeron fan diagrams. The resulting equation would
be, probably, more complicated than our Eq. (\ref{eqN}). It would have
a much more sophisticated kernel \cite{VSF,Cia} and would have cubic
terms in $N$ on the right hand side. However, the analysis presented
in this paper allows one to hope that this resummation will cure some
of the problems of the recently calculated NLO BFKL kernel
\cite{Regge}. Resummation of one loop running coupling corrections,
which are a part of the NLO BFKL kernel, brings a factor of $e^{c
  \alpha^5 Y^3}$ in the one pomeron exchange amplitude, with the
constant $c \approx 5$ for three flavors \cite{Regge}. Since it
introduces an even faster growth with energy than the leading order
BFKL pomeron, this factor seems to complicate the problem of
unitarization of the pomeron. However, the inclusion if this factor of
$e^{c \alpha^5 Y^3}$ in our definition of $P_1$ in Eq. (\ref{n1})
would not change the perturbation series of Eq. (\ref{psol}) and the
quark--antiquark scattering amplitude would still be unitarized by
multiple pomeron exchanges. Of course a careful inclusion of higher
order corrections would lead to much more complicated results.
Nevertheless we may hope that inclusion of one loop running coupling
corrections would just modify the one pomeron exchange contribution
and the multiple pomeron exchanges would still unitarize the total DIS
cross section. However, a rigorous proof of this statement is beyond
our goals here.

Finally we point out that one can use the developed technique to
construct a gluon distribution function $x G (x, Q^2)$ of the nucleus
including multiple hard pomeron exchanges similarly to the way we have
constructed the $F_2$ structure function. To do that one has to
consider scattering of the ``current'' $j = - (1/4) F^a_{\mu \nu}
F^a_{\mu \nu}$ on the nucleus, similarly to
\cite{Mueller4,Mueller6,qc}. Then, in the large $N_c$ limit, it should
be possible to derive an equation for the generating functional of the
dipole wave function of the current $j$, which, at the end, would
provide us with an expression for the gluon distribution function $xG
(x, Q^2)$.

After this paper was finished the author has learned about another
effort to solve Eq. (\ref{eqN}) which was carried out in \cite{lt}.

\section*{Acknowledgments}
I wish to thank Ulrich Heinz, Edmond Iancu and CERN Theory Division
for their hospitality and support during the final stages of this
work.  I would like to thank Gregory Korchemsky, Andrei Leonidov,
Genya Levin, Larry McLerran, Alfred Mueller, Raju Venugopalan and
Samuel Wallon for many helpful and encouraging discussions. This work
is supported in part by DOE grant DE-FG02-87ER40328.

\appendix

\section{}
% Appendix A

Here we are going to show how the Fourier transformation of Eqs.
(\ref{trfm}) and (\ref{invtrfm}) reduces Eq. (\ref{1diff}) to Eq.
(\ref{bsp}). First we note that \cite{Mueller1,GR}
\begin{equation}\label{jac}
  d^2 x_2 = 2 \pi \, x_{12} \, x_{02} \, d x_{12} \, d x_{02} \,
  \int_0^\infty d k \, k \, J_0 (k x_{01}) \, J_0 (k x_{02}) \, J_0 (k
  x_{12}),
\end{equation}
which would allow us to integrate independently over $x_{02}$ and
$x_{12}$ in the kernel of Eq. (\ref{1diff}). Using the Jacobian of Eq.
(\ref{jac}) in the second (quadratic) term on the right hand side of
Eq.  (\ref{1diff}), multiplying both sides of Eq. (\ref{1diff}) by
$J_0 (k x_{01}) / x_{01}$, integrating over $x_{01}$ from $0$ to
$\infty$ and employing
\begin{equation}\label{del}
  \int_0^\infty d x \, x \, J_0 (k x) J_0 (k' x) = \frac{1}{k} \delta
  (k - k')
\end{equation}
we obtain
\begin{eqnarray*}
  \frac{\partial {\tilde N} (k, Y)}{\partial Y} = \frac{\alpha
    N_c}{\pi^2} \int_0^{\infty} \frac{d x_{01}}{x_{01}} J_0 (k x_{01})
  \int_\rho d^2 x_2 \left[ \frac{x^2_{01}}{x^2_{02} x^2_{12}} - 2 \pi
    \delta^2 ({\bf x}_{01} -{\bf x}_{02}) \ln \left(
      \frac{x_{01}}{\rho} \right) \right] N ({\bf x}_{02} , Y) 
\end{eqnarray*}
\begin{eqnarray}\label{2diff}
  - \frac{\alpha N_c}{\pi} {\tilde N}^2 (k, Y).
\end{eqnarray}
The first (linear in $N$) term on the right hand side of Eq.
(\ref{2diff}) can be rewritten employing Eqs. (\ref{jac}),
(\ref{trfm}) and (\ref{invtrfm}) as
\begin{eqnarray}\label{1te}
  \frac{2 \alpha N_c}{\pi} \int_0^{\infty} d k' \, k' \, {\tilde N}
  (k', Y) \int_0^\infty d x_{01} \, x_{01} \, J_0 (k x_{01}) \, J_0
  (k' x_{01}) \, \left( \int_\rho^\infty \, \frac{d x_{12}}{x_{12}} \,
    J_0 (k' x_{12}) - \ln \frac{x_{01}}{\rho}\right).
\end{eqnarray}
Using the techniques outlined in Eqs. (32) and (33) of \cite{Mueller1}
the integral in the parenthesis of Eq. (\ref{1te}) can be done,
yielding
\begin{equation}
  \int_\rho^\infty \, \frac{d x_{12}}{x_{12}} \, J_0 (k' x_{12}) =
  \psi (1) - \ln \frac{k' \rho}{2}.
\end{equation}
Substituting this back into Eq. (\ref{1te}) we end up with
\begin{eqnarray}\label{1te2}
  \frac{2 \alpha N_c}{\pi} \int_0^{\infty} d k' \, k' \, {\tilde N}
  (k', Y) \int_0^\infty d x_{01} \, x_{01} \, J_0 (k x_{01}) \, J_0
  (k' x_{01}) \, \left( \psi (1) - \ln \frac{k' x_{01}}{2}\right).
\end{eqnarray}
Expanding $J_0 (k x_{01})$ in a Taylor series we rewrite Eq.
(\ref{1te2}) as
\begin{eqnarray}\label{1te3}
  \frac{2 \alpha N_c}{\pi} \int_0^{\infty} d k' \, k' \, {\tilde N}
  (k', Y) \sum_{m=0}^{\infty} \frac{(-1)^m}{(m!)^2} \left( \psi (1) -
    \frac{\partial}{ \partial (2 m)} \right) \left[ \left( \frac{k}{2}
    \right)^{2 m} \int_0^\infty d x_{01} \, x_{01} J_0 (k' x_{01}) \,
    x_{01}^{2 m} \right].
\end{eqnarray}
Performing the integration over $x_{01}$ in Eq. (\ref{1te3}) we obtain
[ see Eq. (33) in \cite{Mueller1}]
\begin{eqnarray}\label{1te4}
  \frac{2 \alpha N_c}{\pi} \int_0^{\infty} \frac{d k'}{k'} \, {\tilde
    N} (k', Y) \sum_{m=0}^{\infty} \frac{(-1)^m}{(m!)^2} \left( \psi
    (1) - \frac{\partial}{ \partial (2 m)} \right) \left[ 2 \left(
      \frac{k}{k'} \right)^{2 m} \frac{\Gamma (m + 1)}{\Gamma (- m)}
  \right].
\end{eqnarray}
After differentiating over $m$ and employing the fact that
\begin{eqnarray}\label{del2}
  \sum_{m=0}^{\infty} \frac{(-1)^m}{m!} \left( \frac{k}{k'} \right)^{2
    m} \frac{1}{\Gamma (-m)} = \frac{k}{2} \, \delta (k - k'),
\end{eqnarray}
which is a direct consequence of Eq. (\ref{del}), we can reduce Eq.
(\ref{1te4}) to
\begin{eqnarray}\label{1te5}
  \frac{2 \alpha N_c}{\pi} \int_0^{\infty} \frac{d k'}{k'} \, {\tilde
    N} (k', Y) \sum_{m=0}^{\infty} \frac{(-1)^m}{m!} \, 2 \chi (-2m)
  \left( \frac{k}{k'} \right)^{2 m} \frac{1}{\Gamma (-m)},
\end{eqnarray}
with $\chi$ defined by Eq. (\ref{chi}) above. Rewriting $\chi (-2m)$
in Eq. (\ref{1te5}) as $\chi (- \partial / \partial \ln k)$ and
employing Eq. (\ref{del2}) to perform the summation, which makes the
integration over $k'$ trivial, we obtain the following expression for
the first term on the right hand side of Eq. (\ref{2diff})
\begin{eqnarray}\label{1te6}
  \frac{2 \alpha N_c}{\pi} \chi \left( - \frac{\partial}{\partial \ln
      k} \right) {\tilde N} (k, Y).
\end{eqnarray}
That way Eq.(\ref{2diff}) reduces to Eq. (\ref{bsp}), as desired.

\end{document}